\documentclass[preprint,aps,floats,showpacs]{revtex4}
\usepackage{graphicx}
\usepackage{epsfig}
\usepackage{pstricks}
\usepackage{pst-coil}
\usepackage{amsmath}

\def\PLB#1{{Phys.\ Lett.} {\bf B#1}}

\def\be{\begin{equation}}
\def\ee{\end{equation}}
\def\bea{\begin{eqnarray}}
\def\eea{\end{eqnarray}}

{\newcommand{\lsim}{\mbox{\raisebox{-.6ex}{~$\stackrel{<}{\sim}$~}}}
{\newcommand{\gsim}{\mbox{\raisebox{-.6ex}{~$\stackrel{>}{\sim}$~}}}

\def\ev{\,{\rm eV}}
\def\gev{\,{\rm GeV}}                                                 
\begin{document}
\title{\bf Baryogenesis via leptogenesis in presence of cosmic strings}
\author{Narendra Sahu$^{a,b}$}
\email{narendra@prl.res.in}
\author{Pijushpani Bhattacharjee$^{c,d}$}
\email{pijushb@theory.saha.ernet.in}
\author{Urjit A.~ Yajnik$^a$}
\email{yajnik@phy.iitb.ac.in}
\affiliation{$^a$Department of Physics, Indian Institute of
Technology, Bombay, Mumbai 400076, India}
\affiliation{$^b$Theory Group, Physical Research Laboratory,
Ahmedabad 380 009, India}
\affiliation{$^c$Theory Division, Saha Institute of Nuclear Physics, 
Kolkata-700064, India}
\affiliation{$^d$Indian Institute of Astrophysics, Bangalore-560 034, India}
\pagestyle{empty}

\begin{abstract}
\noindent
We study the effect on leptogenesis due to $B-L$ cosmic strings of  
a $U(1)_{B-L}$ extension of the Standard Model. The disappearance of 
closed loops of $B-L$ cosmic strings can produce heavy right handed 
neutrinos, $N_R$'s, whose $CP$-asymmetric decay in out-of-thermal 
equilibrium condition can give rise to a net lepton ($L$) asymmetry which 
is then converted, due to sphaleron transitions, to a Baryon ($B$) 
asymmetry. This is studied by using the relevant Boltzmann equations and 
including the effects of both thermal and string generated non-thermal 
$N_R$'s. We explore the parameter region spanned by the effective light 
neutrino mass parameter $\tilde{m}_1$, the mass $M_1$ of the lightest of 
the heavy right-handed neutrinos (or equivalently the Yukawa coupling 
$h_1$) and the scale of $B-L$ symmetry breaking, $\eta_{B-L}$, and 
show that there exist ranges of values of these parameters, 
in particular with $\eta_{B-L} > 10^{11}\gev$ and $h_1\gsim 0.01$, for 
which the cosmic string generated non-thermal $N_R$'s can give the 
dominant contribution to, and indeed produce, the observed Baryon 
Asymmetry of the Universe when the purely thermal leptogenesis mechanism 
is not sufficient. We also discuss how, depending on the values of 
$\eta_{B-L}$, $\tilde{m}_1$ and $h_1$, our results lead to 
upper bounds on $\sin\delta$, where $\delta$ is the the $CP$ violating 
phase that determines the $CP$ asymmetry in the decay of the 
heavy right handed neutrino responsible for generating the $L$-asymmetry. 

\end{abstract}
\maketitle

\section{Introduction}
Present low energy neutrino oscillation data~\cite{atmos_data,
solar_data,kamland_data} are elegantly explained by the neutrino 
oscillation hypothesis with very small masses ($\leq$ 1 eV) of 
the light neutrinos. Neutrinos can have either Dirac or Majorana masses. 
Small Majorana masses of the light neutrinos, however, can be 
generated in a natural way through the seesaw mechanism~\cite{seesaw} 
without any fine tuning. This can be achieved by introducing 
right handed neutrinos, $N_R$'s, into the electroweak model 
which are neutral under the known gauge symmetries. The 
Majorana masses of these $N_R$'s are free parameters of the model 
and are expected to be either at TeV scale~\cite{sahu&yajnik_prd.05} 
or at a higher scale~\cite{type-I-bounds,type-II-bounds}. This 
indicates the existence of new physics beyond Standard Model 
(SM) at some predictable high energy scale. 

The heavy right handed Majorana neutrinos are also an essential ingredient 
in currently one of the most favored scenarios of origin of the observed 
baryon asymmetry of the Universe (BAU), namely, the ``baryogenesis via 
leptogenesis" 
scenario~\cite{fukugita.86,luty.92,mohapatra.92,plumacher.96}. 
Majorana mass of the 
neutrino violates lepton number ($L$) and thus provides a natural 
mechanism of generating a lepton asymmetry in the Universe. Specifically,  
leptogenesis can occur via the $L$-violating, $CP$-asymmetric, 
out-of-equilibrium~\cite{sakharov.67} decay of the $N_R$'s into SM leptons 
and Higgs. The resulting $L$-asymmetry is then partially converted to a 
baryon ($B$)-asymmetry via the 
non perturbative $B+L$ violating (but $B-L$ conserving) electroweak 
sphaleron transitions~\cite{krs.85}. The  
attractive aspect of the leptogenesis mechanism is the 
link it implies between the physics of the heavy right handed neutrino 
sector and the experimental data on light neutrino flavor oscillation, 
thus making the scenario subject to experimental tests. Indeed, 
the magnitude of the $L$ (and thus $B$) asymmetry produced depends on, 
among other parameters, the masses of the heavy 
neutrinos, which in turn are related to the light neutrino masses via the 
seesaw mechanism. The mass-square differences amongst the three light 
neutrino species inferred from the results of neutrino experiments, 
therefore, place stringent constraints on the leptogenesis hypothesis. 

A natural way to implement the leptogenesis scenario is to extend
the SM to a gauge group which includes $B-L$ as a gauge charge.
The heavy neutrino masses are then determined by the scale of spontaneous
breaking of this gauge symmetry. Further, with $B-L$ a conserved gauge 
charge and $B+L$ anomalous, we can start with a net $B=L=0$ at a 
sufficiently early stage in the Universe. The observed $B$ asymmetry must 
then be generated only after the phase transition breaking the 
$B-L$ gauge symmetry. 

It is well known that phase transitions associated with spontaneous 
breaking of gauge symmetries 
in the early Universe can, depending on the structure of the symmetry 
group and its breaking pattern, lead to formation of cosmic topological 
defects~\cite{kibble.76,vilen&shell} of various types. In particular, the 
simplest choice for the $B-L$ gauge symmetry group being a 
$U(1)_{B-L}$, the phase transition associated with spontaneous 
breaking of this $U(1)_{B-L}$ in the early Universe would, under very 
general conditions, lead to formation of {\it cosmic 
strings}~\cite{kibble.76,vilen&shell} carrying quantized $B-L$ 
magnetic flux. These ``$B-L$" cosmic strings can be a {\it non-thermal}  
source of $N_R$'s whose decay can give an extra contribution 
to the $L$ and thereby $B$ asymmetry in addition to that from the decay of 
$N_R$'s of purely thermal origin (``thermal'' leptogenesis). This 
can happen in the following two ways: 

First, since the Higgs field defining the $B-L$ cosmic string is the same 
Higgs that also gives mass to the $N_R$ through Yukawa coupling, the 
$N_R$'s can be trapped inside the $B-L$ cosmic strings as fermionic zero 
modes~\cite{jeannerot.96}. Existence of zero energy solutions of fermions  
coupled to a Higgs field that is in a topological vortex string 
configuration is well-known~\cite{jackiw.81,weinberg.81}, and has been 
studied in a variety of models allowing cosmic 
strings~\cite{das.83,witten.85,Stern:1985bg,sdavis.00,vachaspati.01}.
As closed loops of $B-L$ cosmic strings oscillate, 
they lose energy due to emission of gravitational radiation and shrink 
in size. Eventually, when the size of the loop becomes of the order of 
the width of the string, the string loop disappears into massive 
particles among which will be the $N_R$'s which were trapped inside the 
string as zero modes. Each closed loop 
would be expected to release at least one $N_R$ when 
it finally disappears, and the decay of these $N_R$'s would 
then give a contribution to the BAU through 
the leptogenesis route~\cite{jeannerot.96}. 

Second, collapsing, decaying 
or repeatedly self-intersecting closed loops of 
cosmic strings would in general produce heavy gauge and Higgs bosons of 
the underlying spontaneously broken gauge theory. In the context of 
cosmic strings in Grand Unified Theories (GUTs), the $CP$ asymmetric 
$B$-violating decay of the heavy gauge and Higgs bosons released from 
cosmic string loops would produce a net $B$ 
asymmetry~\cite{pijush.82,brandenberger&co.91,riotto&lewis.94,pijush.98}. 
The sphaleron transitions would of course erase the $B$ asymmetry 
{\it unless a net $B-L$ was generated}. If the strings under consideration 
are the $B-L$ cosmic strings, which can be formed at an intermediate stage 
of symmetry breaking in a GUT model based on $SO(10)$, for example, then 
the heavy gauge and Higgs bosons released from the decaying or collapsing 
$B-L$ cosmic string loops can themselves decay to 
$N_R$'s since the $N_R$'s have Yukawa and gauge couplings to the 
(string-forming) Higgs and gauge boson, respectively.  
The decay of these Higgs and gauge boson generated $N_R$'s can produce a 
net $B-L$ and thus contribute to the BAU through the leptogenesis route 
irrespective of the existence of zero modes of $N_R$'s on 
cosmic strings. 

In a previous work~\cite{bha_sahu_yaj.04}, we made a general analytical 
estimate of the contribution of the non-thermal $N_R$'s produced by $B-L$ 
cosmic string loops to BAU. It was shown there that, in order for the 
resulting $B$ asymmetry not to exceed the measured BAU inferred from the 
WMAP results~\cite{spergel.03}, the mass $M_1$ of lightest right handed
Majorana neutrino had to satisfy the constraint 
\be
M_1 \lsim 2.4 \times 10^{12}\left(\eta_{B-L}/10^{13}
\gev\right)^{1/2}\gev,
\label{M1-bound}
\ee
where $\eta_{B-L}$ is the $U(1)_{B-L}$ symmetry breaking scale. In 
the above mentioned analytical study we had (a) taken the $CP$ asymmetry 
parameter $\epsilon_1$ to have its maximum value (see below), (b) not 
taken into account the contribution to BAU from the decay of the already 
existing thermal $N_R$'s, and (c) neglected all wash-out effects (see 
below) on the final $B-L$ asymmetry. Indeed, our aim there was to make a 
simple analytical estimate of the possible maximum contribution of  
the cosmic string generated non-thermal $N_R$'s to the measured BAU. It 
is, of course, clear that a complete analysis of leptogenesis  
in presence of cosmic strings can only be done by solving the 
full Boltzmann equations~\cite{luty.92,plumacher.96} that include the 
non-thermal $N_R$'s of cosmic string origin as well as the already 
existing thermal $N_R$'s and take into account all the relevant 
interaction processes including the wash out effects. This is the study 
taken up in this paper.

The main results obtained in the present paper can be summarized as 
follows: First, we confirm the result, obtained earlier in our analytical 
study~\cite{bha_sahu_yaj.04}, that $B-L$ cosmic string loops can give 
significant contribution to BAU only for $\eta_{B-L}\gsim 10^{11}\gev$. 
Second, the numerical solution of the relevant Boltzmann equations in the 
present paper has enabled us to track the dynamical evolution of the 
contribution of the cosmic string generated non-thermal $N_R$'s to the 
final BAU. Specifically, we find that for sufficiently large values of 
$\eta_{B-L}$ and for appropriate ranges of values of the other 
relevant parameters ${\tilde m_1}$ and $h_1$ as detailed in 
sec.~\ref{subsec:solving}, the effect of the cosmic string generated 
non-thermal $N_R$'s is to produce a late-time increase of the final 
value of BAU as compared to its value in absence of cosmic strings.    
This can be understood from the fact that while the thermal abundance of 
$N_R$'s decreases exponentially with decreasing temperature, the density 
of cosmic string generated $N_R$'s has a power law dependence (on 
temperature) inherited from the scaling behavior 
of the evolution of cosmic strings, leading to a domination of the 
string generated $N_R$'s over the thermal $N_R$'s at late times for 
sufficiently large values of $\eta_{B-L}$. 
In such situations, we are required to place an {\it upper} bound on the 
magnitude of $\sin\delta$, where $\delta$ is the $CP$ violating phase that 
determines the $CP$ asymmetry in the heavy neutrino sector, in 
order not to overproduce the BAU.  

The rest of our paper is organized as follows. In section \ref{sec:brief}, 
we briefly review the standard thermal baryogenesis via leptogenesis 
scenario, and discuss the required lower bounds on the mass of the 
lightest right handed neutrino and the $B-L$ symmetry breaking scale. 
In section \ref{sec:cosmic} we introduce closed loops of $B-L$ cosmic 
strings as non-thermal sources of $N_R$'s and write 
down the injection rates of these $N_R$'s due to the two main 
processes of disappearance of cosmic string loops. Section 
\ref{sec:complete} is devoted to setting up and then solving the relevant 
Boltzmann equations for the evolution of the $B-L$ asymmetry, including 
the non-thermal $N_R$'s of cosmic string origin in addition to the usual 
thermal ones. The effects of the cosmic string generated non-thermal 
$N_R$'s on the evolution of the $B-L$ asymmetry are discussed. Finally, 
section \ref{sec:conclusion} summarizes our main results. 

\section{A brief review of baryogenesis via leptogenesis
\label{sec:brief}}

\subsection{The general framework\label{subsec:general}}
In the model we consider, the lepton asymmetry arises through the 
decay of right handed Majorana neutrinos to the SM 
leptons ($\ell$) and Higgs ($\phi$) via the Yukawa 
coupling
\begin{equation}
\mathcal{L}_{Y}=f_{ij}\bar{\ell}_i\phi N_{R_j}+{\rm h.c.}\,,
\label{yukawa_term}
\end{equation}
where $f_{ij}$ is the Yukawa coupling matrix, and $i,j=1,2,3$
for three flavors. 

We use the basis in which the mass matrix of the heavy Majorana
neutrinos, $M$, is diagonal,  where the Majorana neutrinos are defined 
by $N_j=\frac{1}{\sqrt{2}}(N_{R_j}+N^c_{R_j})$. However, in 
this basis the Dirac mass matrix $m_D$ ($=f v$, $v$ being the 
vacuum expectation value of the SM Higgs) of the neutrinos 
is not diagonal. The canonical see-saw mechanism then gives the 
corresponding light neutrino mass matrix 
\be
m_\nu=-m_DM^{-1}m_D^T.
\label{neu-mass-matrix}
\ee
In this basis the light neutrino mass matrix $m_{\nu}$ can
be diagonalized by the lepton mixing matrix~\cite{mns-matrix}
$U_L$ to give the three light neutrino masses, 
\be
{\rm diag}(m_1, m_2, m_3)=U_L^T m_{\nu}U_L. 
\label{dia-mass-matrix}
\ee

The best fit values of the mass-square differences 
and mixing angles of the light neutrinos from a global three neutrino 
flavors oscillation analysis are~\cite{gonzalez-garcia_prd.03}
\be
\theta_{\odot}\equiv \theta_{1}\simeq 34^\circ, 
~~\theta_{atm}\equiv\theta_{2}=45^\circ, ~~\theta_3 \leq 13^\circ,
\label{bestfit-theta}
\ee
and
\begin{eqnarray}
\Delta m_{\odot}^2\equiv |m_2^2 - m_1^2| &\simeq& 7.1\times 10^{-5}
\ev^2\,, \label{sol-masses}\\
\Delta m_{atm}^2\equiv |m_3^2 - m_1^2| &\simeq & 2.6\times 10^{-3}
\ev^2\,.
\label{atm-masses}
\end{eqnarray}
In the absence of data about the overall scale of neutrino masses, we 
shall assume throughout this paper a normal hierarchy,
$m_1\ll m_2\ll m_3$, for the light neutrino masses.
With this assumption equations (\ref{sol-masses}) and (\ref{atm-masses}) 
give 
\bea
m_2\simeq \sqrt{\Delta m_{\odot}^2}&=& 0.008 \ev\,,\nonumber\\
m_3\simeq \sqrt{\Delta m_{atm}^2}&=&0.05\ev\,.
\label{bestfit-masses}
\eea
We will use the above value of $m_3$ to obtain the upper 
bound on $CP$-asymmetry in subsection \ref{subsec:decay}.

\subsection{Decay of heavy Majorana neutrino and $CP$-asymmetry
\label{subsec:decay}}
The $CP$-asymmetry parameter in the decay of $N_i$ is defined as
\be
\epsilon_{i}\equiv\frac{\Gamma(N_i\rightarrow \bar{\ell}\phi)-
\Gamma(N_i\rightarrow \ell\phi^\dagger)}{\Gamma(N_{i}
\rightarrow \bar{\ell}\phi)+\Gamma(N_{i}\rightarrow \ell\phi^\dagger)}\,.
\label{epsilon_def}
\ee
We assume a normal mass hierarchy in the heavy Majorana
neutrino sector, $M_{1} < M_{2} < M_{3}$, and further assume that the 
final lepton asymmetry is produced mainly by the decays of the lightest 
right handed Majorana neutrino, $N_{1}$. The latter is 
justified~\cite{luty.92,plumacher.96} because any asymmetry 
produced by the decay of $N_{2}$ and $N_{3}$ is erased by the lepton 
number violating interactions mediated by $N_{1}$. At an epoch of 
temperature $T>M_1$, all the lepton number violating processes 
mediated by $N_1$ are in thermal equilibrium. As the Universe 
expands and cools to $T\lsim M_1$, the $L$-violating scatterings mediated
by $N_{1}$ freeze out, thus providing the out-of-equilibrium
situation~\cite{sakharov.67} necessary for the survival of any 
net $L$-asymmetry generated by the decay of the $N_1$'s. The 
final $L$-asymmetry is, therefore, given essentially by the 
product of the equilibrium 
number density of the $N_1$'s at $T\sim M_1$ and the 
$CP$-asymmetry parameter $\epsilon_1$. The latter is 
given by~\cite{type-I-bounds,type-II-bounds}
\be
|\epsilon_{1}|=\frac{3}{16\pi v^2}M_1 m_3 \sin\delta\,,
\label{epsilon_1}
\ee
where $v\simeq 174\gev$ is the electroweak symmetry breaking
scale and $\delta$ is the $CP$ violating phase. Using the best 
fit value of $m_3$ from equation (\ref{bestfit-masses}) the value of
$\epsilon_{1}$ can be written as
\be
|\epsilon_{1}|\leq 9.86\times 10^{-8}\left(\frac{M_1}{10^9\gev}
\right)\left(\frac{m_3}{0.05\ev}\right)\,.
\label{epsilon1-max}
\ee

\subsection{Thermal leptogenesis and bound on $M_1$
\label{subsec:baryogenesis}}
The $L$-asymmetry, created by the decays of $N_1$, is
partially converted to a $B$-asymmetry by the 
nonperturbative sphaleron transitions which violate
$B+L$ but preserve $B-L$. Assuming that sphaleron
transitions are ineffective at temperatures below 
the electroweak transition temperature ($T_{\rm EW}$),
the $B$-asymmetry is related to $L$-asymmetry by the
relation~\cite{harvey_turner}
\be
B=\frac{p}{p-1}L\simeq -0.55 L\,,
\label{B_L_relation}
\ee
where we have taken $p=28/79$ appropriate for the particle
content in SM. The net baryon asymmetry of the Universe, defined as 
$Y_B\equiv (n_B/s)$, can thus be written as 
\be
Y_B\simeq 0.55\epsilon_1 Y_{N_1}|_{T\simeq M_1}\,d\,,
\label{bau_def}
\ee
where the factor $0.55$ in front comes from equation (\ref{B_L_relation}), 
\be
Y_{N_1}\equiv\frac{n_{N_1}}{s}\,,
\label{com-neu-den}
\ee
$n_{N_1}$ being the number density of $N_1$, 
and $d$ is the dilution factor due to wash-out effects. In the 
above equations $s$ stands for the entropy density and is given by 
\be
s=\frac{2\pi^2}{45}g_{*}T^{3}\simeq 43.86 (g_*/100)T^3\,,
\label{entropy}
\ee
where $g_*$ is the total number of relativistic degrees of freedom 
contributing to entropy of the Universe. 
  
The present-day observed value of the baryon-to-photon ratio $\eta\equiv
(n_B-n_{\bar{B}})/n_\gamma$ inferred from the WMAP data 
is~\cite{spergel.03}
\bea
\eta_0^{\rm WMAP} &\simeq& 7.0\, Y_{B,0}\\
&=&\left(6.1^{+0.3}_{-0.2}\right)\times 10^{-10}\,.
\label{eta_0_wmap}
\eea 
Using equation (\ref{epsilon1-max}) in equation (\ref{bau_def}) and 
comparing with (\ref{eta_0_wmap}) we get   
\be
M_1\geq 10^9\gev \times\left(\frac{1.6\times 
10^{-3}}{Y_{N_1}|_{T\simeq M_1}d }
\right)\left(\frac{0.05 eV}{m_3}\right)\,.
\label{lower-bound-M1}
\ee
Assuming that $N_1$'s are initially (i.e., at $T\gg M_1$) in thermal 
equilibrium and they remain so till $T\simeq M_1$, one has  
$Y_{N_1}|_{T\simeq M_1}\approx 2.3\times10^{-3}\times (100/g_*)$.  
Equation (\ref{lower-bound-M1}) then indicates that the $B-L$ symmetry 
breaking scale must satisfy the constraint $\eta_{B-L}\gsim  
O(10^9)\gev$ for successful thermal leptogenesis, unless we allow the 
Majorana neutrino Yukawa coupling of the lightest right handed neutrino, 
$h_1$, to be greater than unity. 

\section{$B-L$ cosmic strings as sources of non-thermal heavy 
neutrinos\label{sec:cosmic}}
\subsection{$B-L$ cosmic strings and zero modes of $N_R$ 
\label{subsec:zeromodes}}
The possibility of $B-L$ cosmic strings in various extensions
of the SM were studied in ~\cite{jeannerot&co, yajnik.99}. 
For simplicity, we consider in the present paper a model based on the 
gauge group 
$SU(2)_{\rm L}\otimes U(1)_{\rm Y}\otimes U(1)_{\rm Y'}$, where 
$Y'$ is a linear combination of $Y$ and $B-L$~\cite{buc_gre_min.91}. 
We then follow the symmetry breaking scheme
\begin{eqnarray}
SU(2)_{L}\otimes U(1)_{Y} \otimes U(1)_{Y'} &\underrightarrow 
{\langle \chi \rangle =\eta_{B-L}}& SU(2)_{L}\otimes 
U(1)_{Y}\nonumber\\
&\underrightarrow{\langle \phi \rangle =v}&  U(1)_{EM},
\label{symmetry_break}
\end{eqnarray}
where $\chi$ is the Higgs boson required to break the 
$U(1)_{Y'}$ gauge symmetry and $\phi^{T}=(\phi^{+}, \phi^{0})$ 
is the SM Higgs. 
 
As the Universe cools below the critical temperature, 
$T_{B-L}$, of the $B-L$ symmetry breaking phase 
transition, the Higgs field $\chi$ develops a vacuum 
expectation value $\langle \chi \rangle =\eta_{B-L}$. The same
Higgs field also forms strings. The mass per unit length of the string, 
$\mu$, is given by~\cite{vilen&shell} $\mu\sim\eta^{2}_{B-L}\sim 
T_{B-L}^2$. The exact value of $\mu$ depends on the values of the 
parameters of the model, in particular the relevant gauge and Higgs boson 
masses, and can differ from $\eta_{B-L}^2$ by up to an order of 
magnitude depending on the model~\cite{hill-hodges-turner-prl.87}. 
Measurements of CMB anisotropies have been used to place upper bounds on 
the fundamental cosmic string parameter $\mu$ in a variety of different 
cosmic string models; see, for example, \cite{cmb_limits}. 
A recent analysis~\cite{wyman_etal} of data on CMB anisotropies and large 
scale structure together puts the bound $G\mu < 3.4 (5)\times 10^{-7}$ 
at 68 (95)\% c.l. This most likely rules out cosmic string formation at a 
typical GUT scale $\sim10^{16}\gev$. However, lighter cosmic strings 
arising from symmetry breaking at lower scales, for example, the $B-L$ 
cosmic strings discussed in this paper, are not ruled out.  

The right handed neutrinos acquire Majorana masses from 
coupling to the same Higgs field $\chi$, 
\be
-\mathcal{L}_{\chi-{N_R}}=\frac{1}{2}[ih_{ii}\chi\overline{N_{R_{i}}}
N_{R_{i}}^{c}+{\rm h.c}]\,,
\label{nu-chi-int}
\ee
where $h$ is the Yukawa coupling matrix, and $N_{R}^{c}
=i \sigma^{2}N_{R}^{*}$ defines the Dirac charge conjugation 
operation. The equation of motion of $N_R$ in the background of 
$\chi$ forming the string admits $|n|$ normalizable zero-modes 
in the winding number sector 
$n$~\cite{jackiw.81,weinberg.81,jeannerot.96,sdavis.00}. On a straight 
string these modes are massless. 
However, on a generic string they are expected to acquire effective 
masses proportional to the inverse radius of the string
curvature. As soon as this mass becomes comparable to the 
mass of the free neutrinos in the bulk medium, these neutrinos 
can be emitted from the string. In any
case, when the string loop shrinks and finally decays it emits 
various massive particles: the gauge bosons, the heavy Higgs ($\chi$) 
and the massive right handed Majorana neutrinos ($N_i$). 

\subsection{$N_R$'s from closed loops of $B-L$ cosmic strings
\label{subsec:N_R_from_closed_loops}}
A key physical process that governs the evolution of cosmic string 
networks in an expanding Universe is the formation of sub-horizon sized
closed loops which are pinched off from the network whenever a
string segment curves over into a loop and intersects itself. It is this 
process that allows the string network to reach a scaling regime, in which 
the energy density of the string network scales as a fixed fraction of the 
radiation or matter energy density in the Universe; see, e.g.,
Ref.~\cite{vilen&shell} for a text book
discussion of evolution of cosmic strings in the Universe. 

After their formation, the closed loops eventually disappear through 
either of the following two processes (see 
Refs.~\cite{bha_sahu_yaj.04,pijushreport} for the relevant details):  

\subsubsection{Slow death\label{subsubsec:slow}}
Closed loops born in non-selfintersecting configurations oscillate freely 
with oscillation time period $L/2$ ($L$ being the length of the loop). In 
doing so they slowly lose energy due to emission of gravitational 
radiation, and thus shrink in size. Eventually, when the radius of the 
loop becomes of the order of the width ($\sim\eta_{B-L}^{-1}$) of the 
string --- which happens over a time scale large compared to $L$ (hence 
``slow'') --- the resulting ``tiny'' loop loses 
topological stability and decays into elementary particle quanta including  
the gauge- and Higgs bosons associated with the underlying broken symmetry 
as well the heavy neutrinos $N_i$'s coupled to the gauge- and Higgs  
bosons. While we expect the final number $N_N$ of the heavy  
neutrinos released per tiny loop to be of order one, it is difficult 
to be more precise, and we keep this number as an undetermined parameter. 
Assuming that the energy of a loop going into $N_R$'s is predominantly in 
the form of the lightest of the heavy neutrinos, $N_1$'s, the 
number of heavy Majorana neutrinos released from the closed loops 
disappearing through this ``slow death'' (SD) process per unit time per 
unit volume at any time $t$ (in the radiation dominated epoch) can be 
written as~\cite{bha_sahu_yaj.04}
\be
\left(\frac{dn_{N_1}}{dt}\right)_{\rm SD} = 
N_N f_{\rm SD}\frac{1}{x^{2}}\left(\Gamma G \mu\right)^{-1}
\frac{(K+1)^{3/2}}{K}t^{-4}\,,
\label{N_SD_rate}
\ee
where $f_{\rm SD}$ is the fraction of newly born loops which
die through the SD process, $x\sim0.5$ is a numerical factor that 
characterizes the scaling configuration of the string network, $\Gamma\sim 
100$ is a numerical factor that determines the life time of a 
loop due to gravitational radiation emission, and $K\sim O(1)$ is a 
numerical factor that determines the average length of the closed loops 
at their birth. It is generally argued~\cite{vilen&shell} that $f_{\rm 
SD}\simeq 1$.

Using (\ref{com-neu-den}), the above rate then gives the injection rate of 
massive Majorana neutrinos from cosmic string closed loops disappearing 
through the SD process per comoving volume as 
\be
\left(\frac{dY^{\rm st}_{N_1}}{dz}\right)_{\rm SD}= \frac{1.57\times 
10^{-17}}{z^4}
N_N \left(\frac{M_1}{\eta_{B-L}}\right)^2
\left(\frac{M_1}{\gev}\right),
\label{slowdeath-rate}
\ee
where $z=M_1/T$ is the dimensionless variable with respect to
which the evolution of the various quantities is studied. In the above  
we have used the following numerical values for the various constants:  
$\Gamma=100$, $x=0.5$, $g_{*}=100$ and $K=1$. 

\subsubsection{Quick death\label{subsubsec:quick}}
Some fraction of closed loops may be born in configurations with
waves of high harmonic number. Such string loops have been
shown~\cite{siemens&kibble.95} to have a high probability of
self-intersecting. In this case a loop of length $L$
can break up into a debris of tiny loops of size $\eta_{B-L}
^{-1}$ (at which point they turn into the constituent massive 
particles) on a time-scale $\sim L$ (hence ``quick''). Since gravitational 
energy loss occurs over a time scale much larger than $L$, these loops do 
not radiate any significant amount of energy in gravitational radiation, 
and thus almost the entire original energy of these loops would eventually 
come out in the form of massive particles.

Assuming again that each tiny loop of size 
$\sim \eta_{B-L}^{-1}$ yields a number $N_N$ of heavy neutrinos, 
the injection rate of the massive Majorana neutrinos due to this ``quick 
death'' (QD) process is given by~\cite{bha_sahu_yaj.04}
\be
\left(\frac{dn_{N_1}}{dt}\right)_{\rm QD}=f_{\rm QD} N_N \frac{1}{x^{2}}
\frac{\eta_{B-L}}{t^3}\,, 
\label{quick_N_rate}
\ee 
where $f_{\rm QD}$ is the fraction of loops that undergo QD. 

Note that while each tiny loop yields the same number $N_N$ of the 
heavy neutrinos irrespective of the SD or QD nature of the demise of the 
loop, the total injection rates of the heavy neutrinos in the two cases 
are different because of the different number of tiny loops that result  
from an initial big loop and the different time scales involved in the two 
cases. 

Using (\ref{com-neu-den}) the above rate 
(\ref{quick_N_rate}) can be rewritten as 
\be
\left(\frac{dY^{\rm st}_{N_1}}{dz}\right)_{\rm QD}\simeq \frac{1.36\times 
10^{-36}}{z^2} f_{\rm QD} N_N \left(\frac{\eta_{B-L}}{\gev}\right)
\left(\frac{M_1}{\gev}\right).
\label{quickdeath-rate}
\ee

While the value of $f_{\rm QD}$ is not known, there are 
constraints~\cite{pijush&rana,pijushreport} on $f_{\rm QD}$ from the 
measured flux of the ultrahigh energy cosmic rays above $\sim 
5\times10^{19}\ev$ and more stringently from the 
cosmic gamma ray background (CGRB) in the 10 MeV -- 100 GeV energy
region measured by the EGRET experiment~\cite{sreekumar}. The latter gives
\be
f_{\rm QD}\eta_{16}^{2}\lsim 9.6\times 10^{-6}\,,
\label{f_QD_constraint}
\ee
where $\eta_{16}\equiv(\eta_{B-L}/10^{16}\gev)$. For GUT
scale cosmic strings with $\eta_{16}=1$, for example, the above
constraint implies that $f_{\rm QD}\leq 10^{-5}$, so that
most loops should be in non-selfintersecting configurations,
consistent with our assumption of $f_{\rm SD}\sim 1$.
Note, however, that $f_{\rm QD}$ is not constrained by
(\ref{f_QD_constraint}) for cosmic strings formed at 
scales $\eta_{B-L}\lsim 3.1\times 10^{13}\gev$.

\section{Leptogenesis in presence of cosmic strings: towards a complete 
analysis\label{sec:complete}}
\subsection{Analytical estimate\label{subsec:analytical}}
\begin{figure}[ht]
\epsfig{file=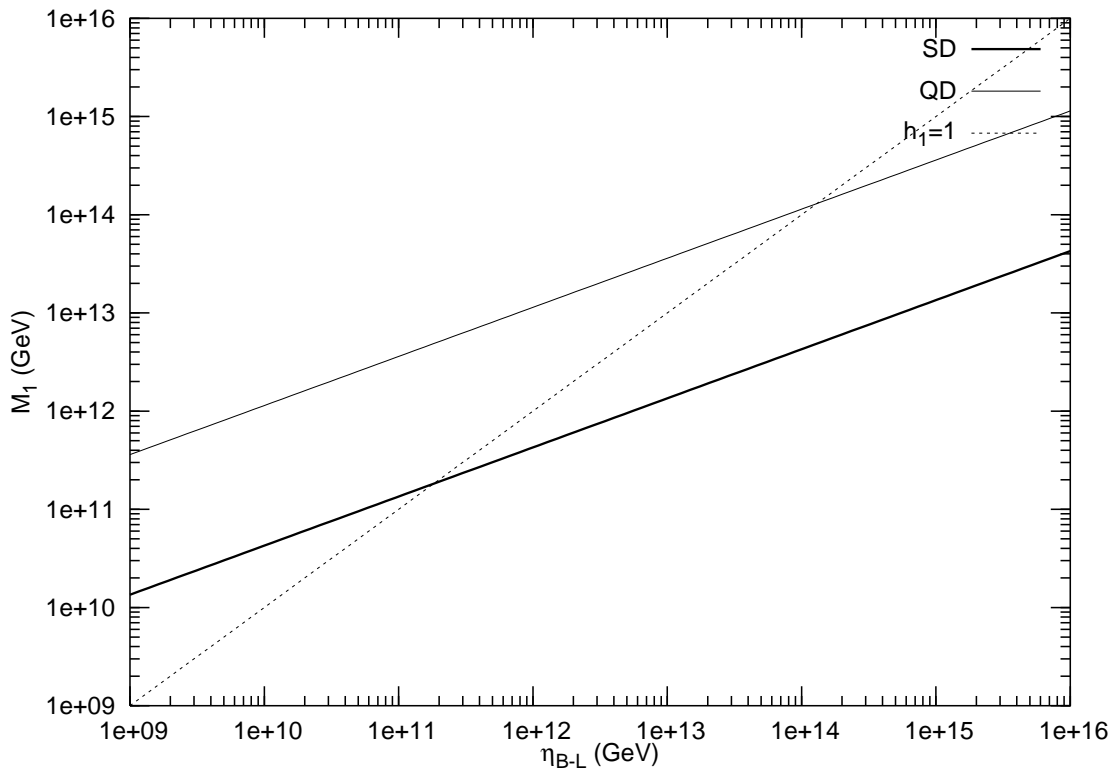,angle=-90,width=0.9\textwidth}
\caption{Constraints on $h_1=M_1/\eta_{B-L}$ from consideration of 
possible maximal contributions of $B-L$ cosmic string loops to 
$B$-asymmetry. Models lying on the thick (thin) solid line can in 
principle produce the observed BAU entirely due to the SD (QD) process of 
$B-L$ cosmic string loops. Models above the thick solid line are ruled 
out from consideration of overproduction of the $B$ asymmetry. The 
$h_1=1$ line is also shown for comparison.}
\label{fig:region}
\end{figure}

In our previous work~\cite{bha_sahu_yaj.04} we made simple analytical 
estimates of the maximum possible contributions of cosmic string loops to 
leptogenesis. Neglecting the contribution of the thermal $N_R$'s, allowing 
the maximal value of the $CP$-violation parameter given by equation 
(\ref{epsilon1-max}) and demanding that the resulting value of the 
baryon-to-photon ratio not exceed the observed value given by equation 
(\ref{eta_0_wmap}), we derived upper bounds on the mass $M_1$ for the SD 
and QD processes of decay of cosmic string loops. This is shown 
in Figure ~\ref{fig:region} for $N_N=10$ for illustration. 

Figure \ref{fig:region} allows us to identify regions in the 
$\eta_{B-L}-M_1$ plane determining models for which cosmic string 
processes discussed above can play a significant role in the leptogenesis 
mechanism. Clearly, from Figure \ref{fig:region}, cosmic string are 
relevant for leptogenesis  
only for $\eta_{B-L}\gsim 1.7\times10^{11}\gev$; lower values of 
$\eta_{B-L}$ are relevant only if we allow $h_1\geq1$. 
Note also that the above lower limit on $\eta_{B-L}$ (for cosmic strings  
to be relevant for leptogenesis) is determined by the SD process; the QD 
process becomes relevant only at much higher values of $\eta_{B-L}$, 
namely, $\eta_{B-L}^{\rm QD}\gsim 1.2\times 10^{14}\gev$. 

In this context, it may be mentioned here that, following our previous 
work, a recent work~\cite{gu-mao-plb.05} has found that in the case of 
degenerate neutrinos (as opposed to hierarchical neutrinos assumed in our 
work), the $B-L$ cosmic strings become relevant for 
leptogenesis at much higher values of $\eta_{B-L}$ compared to those in  
the case of hierarchical neutrinos, e.g., $\eta_{B-L}\gsim 
3.3\times10^{15}\gev$ for the SD process and $\eta_{B-L}>10^{16}\gev$ for 
the QD process. As already mentioned, we assume hierarchical neutrino 
masses in the present paper. 

We now proceed towards a complete analysis of leptogenesis in presence 
of cosmic strings by first setting up and then solving the relevant 
Boltzmann equations that include the non-thermal $N_R$'s 
produced from the decaying cosmic string loops as well as the 
already existing thermal $N_R$'s and also include all the 
relevant wash-out effects. 

\subsection{Boltzmann Equations}
The total rate of change of the abundance of $N_1$'s can be written as 
\be
\frac{dY_{N_1}}{dz}=\left(\frac{dY_{N_1}}{dz}\right)_{D,S}+
\left(\frac{dY_{N_1}}{dz}\right)_{\rm injection}.
\label{eff-rate}
\ee
The first term on the right hand side of equation (\ref{eff-rate}) 
is given by the usual Boltzmann equation ~\cite{luty.92,plumacher.96} 
\be
\left(\frac{dY_{N_1}}{dz}\right)_{D,S}=
-(D+S)\left(Y_{N_1}-Y^{\rm eq}_{N_1}\right)\,,
\label{thermal-rate}
\ee
where $D$ and $S$ constitute the decay and $\Delta L=1$ 
lepton number violating scatterings, respectively, which reduce the number 
density of $N_1$, and $Y^{\rm eq}_{N_1}$ is the abundance of $N_1$ 
in thermal equilibrium. 

The second term on the right hand side of equation (\ref{eff-rate}) gives 
the rate of injection of $N_1$'s due to disappearance of $B-L$ 
cosmic string loops:  
\be
\left(\frac{dY_{N_1}}{dz}\right)_{\rm injection}=\left(
\frac{dY_{N_1}^{\rm st}}{dz}\right)_{\rm SD} + \left(\frac{dY_{N_1}^{\rm 
st}}{dz}\right)_{\rm QD},
\label{nonthermal-rate}
\ee
where the two terms on the right hand side are given by 
the equations (\ref{slowdeath-rate}) and (\ref{quickdeath-rate}),  
respectively. 

The two terms in equation (\ref{eff-rate}) compete with 
each other. While the first term reduces the density 
of $N_1$'s, the second term tries to increase it 
through continuous injection of $N_1$'s from 
the disappearance of cosmic string loops. The CP-violating decays  
of the $N_1$'s produce a net $B-L$ asymmetry. This can be 
calculated by solving the Boltzmann equation 
\be
\frac{dY_{B-L}}{dz} = -\epsilon_{1} D\left(Y_{N_1}-Y^{\rm eq}_
{N_1}\right)-W Y_{B-L}.
\label{asymmetry-rate}
\ee
The first term in equation (\ref{asymmetry-rate})  
involving the decay term $D$ produces an asymmetry while a part 
of it gets erased by the wash out terms represented by $W$ which includes 
the processes of inverse decay as well as the $\Delta L=1$ and  
$\Delta L=2$ lepton number violating scatterings. In a recent 
work~\cite{strumia.04} it has been claimed that thermal 
corrections to the above processes as well as processes 
involving gauge bosons are important for the final $B-L$ asymmetry. 
However, this is under debate~\cite{buch-bari-plum-ped.04}, and we have 
not included these corrections in the present paper.  

In equations (\ref{thermal-rate}) and (\ref{asymmetry-rate}),  
$D=\Gamma_D/Hz$, where  
\be
\Gamma_D=\frac{1}{16 \pi v^2}\tilde{m}_1 M_1^2
\label{decay}
\ee
is the decay rate of $N_1$, with 
\be
\tilde{m}_1 \equiv \frac{(m_D^\dagger m_D)_{11}}{M_1} 
\label{eff-neu-mass}
\ee
the {\it effective neutrino mass} parameter~\cite{plumacher.96}, 
and $H$ is the Hubble expansion parameter. Also, 
$S=\Gamma_S/Hz$, where
$\Gamma_S$ is the rate of $\Delta L=1$ lepton number 
violating scatterings and $W=\Gamma_W/Hz$, where $\Gamma_W$ 
is the rate of wash out effects involving the $\Delta L=1$ and  
$\Delta L=2$ lepton number violating processes and 
inverse decay. The various $\Gamma$s are related to the scattering 
densities~\cite{luty.92} $\gamma$'s as
\begin{equation}
\Gamma_i^X(z)=\frac{\gamma_i(z)}{n_X^{\rm eq}}.
\end{equation}
The dependence of the scattering rates involved in $\Delta L=1$ 
lepton number violating processes on $\tilde{m}_1$ and $M_1$ are 
similar to that of the decay rate $\Gamma_{D}$. As the 
Universe expands these $\Gamma$'s compete with the Hubble 
expansion parameter. In a comoving volume we have (with  
same notations as in ~\cite{luty.92}) 
\be
\left(\frac{\gamma_{D}}{sH(M_1)}\right), \left(\frac{
\gamma^{N_1}_{\phi,s}}{sH(M_1)}\right), \left(\frac{
\gamma^{N_1}_{\phi,t}}{sH(M_1)}\right) \propto k_1\tilde{m}_1.
\label{dilution}
\ee
On the other hand the $\gamma$'s for the 
$\Delta L=2$ lepton number violating processes depend on
$\tilde{m}_1$ and $M_1$ as 
\be
\left(\frac{\gamma^l_{N_1}}{sH(M_1)}\right), \left(\frac{
\gamma^l_{N_1,t}}{sH(M_1)}\right) \propto k_2 \tilde{m}_1^2 M_1.
\label{washout}
\ee  
In the above equations (\ref{dilution}), (\ref{washout}), 
$k_i$, $i=1,2,3$ are dimensionful constants determined from 
other parameters; see ~\cite{luty.92} for details. 

\subsection{Constraint on the effective neutrino mass ($\tilde{m}_1$)
\label{subsec:constraint}}
In order to satisfy the out of equilibrium condition the decay rate of 
$N_1$ has to be less than the Hubble expansion parameter. This imposes a 
constraint on the effective neutrino mass parameter $\tilde{m}_1$ as 
follows. At an epoch $T<M_1$,
\be
\frac{\Gamma_D}{H}\equiv \frac
{\tilde{m}_1}{m_*}=K < 1,
\label{decay-con}
\ee  
where $m_*$, the {\it cosmological neutrino mass parameter}, is given 
by~\cite{fglp}
\be
m_*\equiv 4\pi g_*^{1/2}\frac{G_N^{1/2}}{\sqrt{2}G_F}
= 6.4\times 10^{-4}\ev,
\label{crit-mass}
\ee
where $g_*$ is the effective number of relativistic degrees of freedom. 
We see that a net $B-L$ asymmetry can be generated dynamically 
provided $\tilde{m}_1 < m_*$ at an epoch 
$T<M_1$. This constraint on $\tilde{m}_1$ can be realized in a model as 
follows. 

We assume a charge-neutral lepton symmetry for which we take the
texture of the Dirac mass of the neutrino of the 
form~\cite{falcone.00}
\be
m_D=\begin{pmatrix}
0 & \sqrt{m_e m_{\mu}} & 0\\
\sqrt{m_e m_{\mu}} & m_{\mu} & \sqrt{m_e m_{\tau}}\\
0 & \sqrt{m_e m_{\tau}} & m_{\tau}
\end{pmatrix}\,.
\label{dirac-texture}
\ee
Using (\ref{lower-bound-M1}) and (\ref{dirac-texture}) in 
equation (\ref{eff-neu-mass}) we get the constraint to be 
\be
\tilde{m}_1 \leq 5.25\times 10^{-6}\ev, 
\ee    
which is in consonance with the requirement that 
$\tilde{m}_1 < m_*$ at any epoch $T<M_1$.

\subsection{Solving the Boltzmann equations\label{subsec:solving}}
At an epoch $T\gg M_1$ the lepton number violating processes are 
sufficiently fast as to set the $B-L$ asymmetry to zero. As the 
temperature falls and 
becomes comparable to $M_1$, a net $B-L$ asymmetry is generated through 
the $CP$-violating decay of $N_1$. The resulting asymmetry can be 
obtained by solving the Boltzmann equations. We solve equations 
(\ref{eff-rate}) and (\ref{asymmetry-rate}) numerically with the 
following initial conditions
\be
Y^{\rm in}_{N1}=Y_{N_1}^{\rm eq}~~{\mathrm and}~~ Y^{\rm in}_{B-L}=0.
\label{in_condn}
\ee
Using the first initial condition we solve equation 
(\ref{eff-rate}) for $Y_{N_1}$, and the corresponding $B-L$ asymmetry 
$Y_{B-L}$ is obtained from equation (\ref{asymmetry-rate}) by using 
the second initial condition of equation (\ref{in_condn}). 
\begin{figure}[htbp]
\epsfig{file=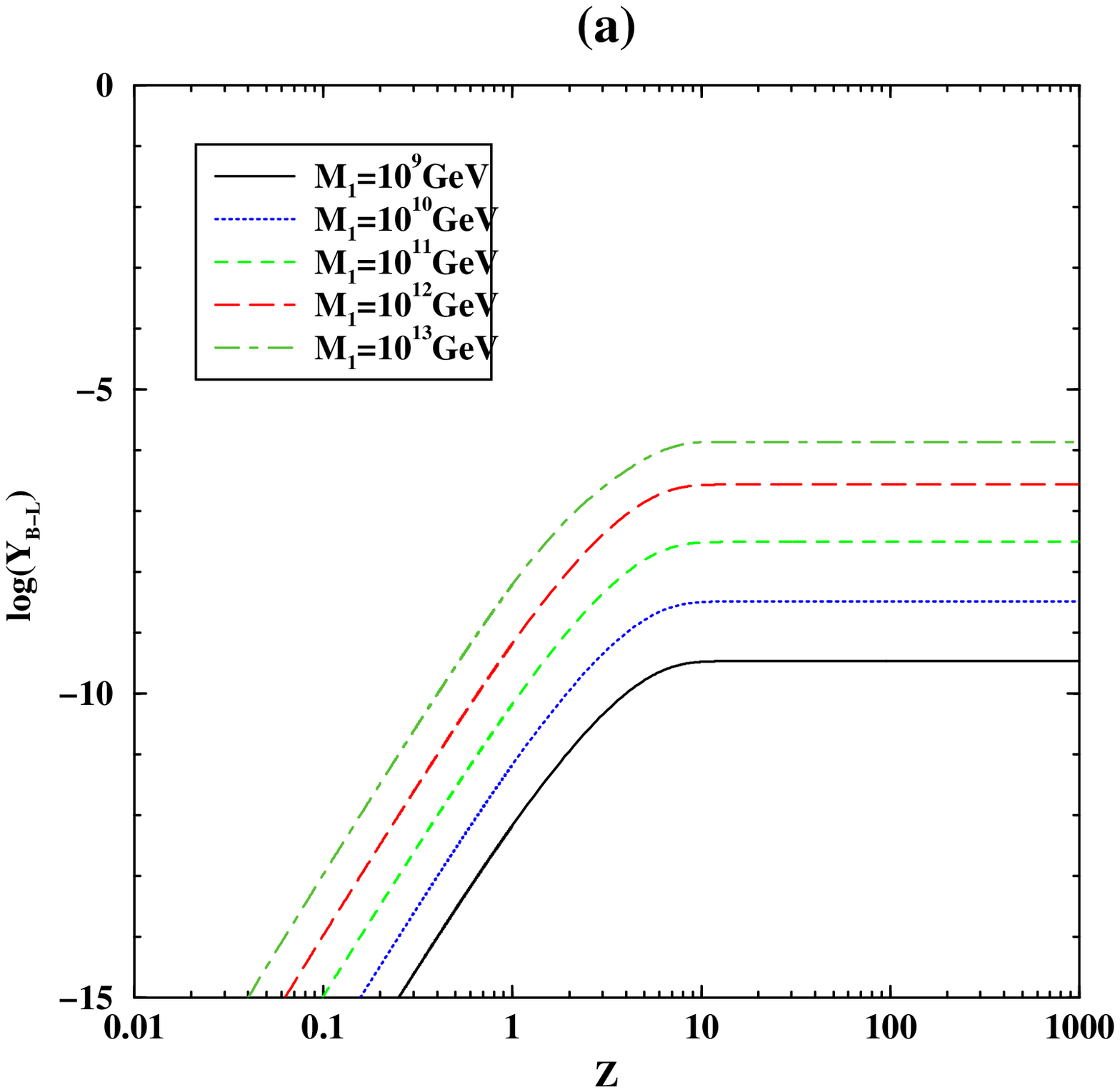, width=0.5\textwidth}
\epsfig{file=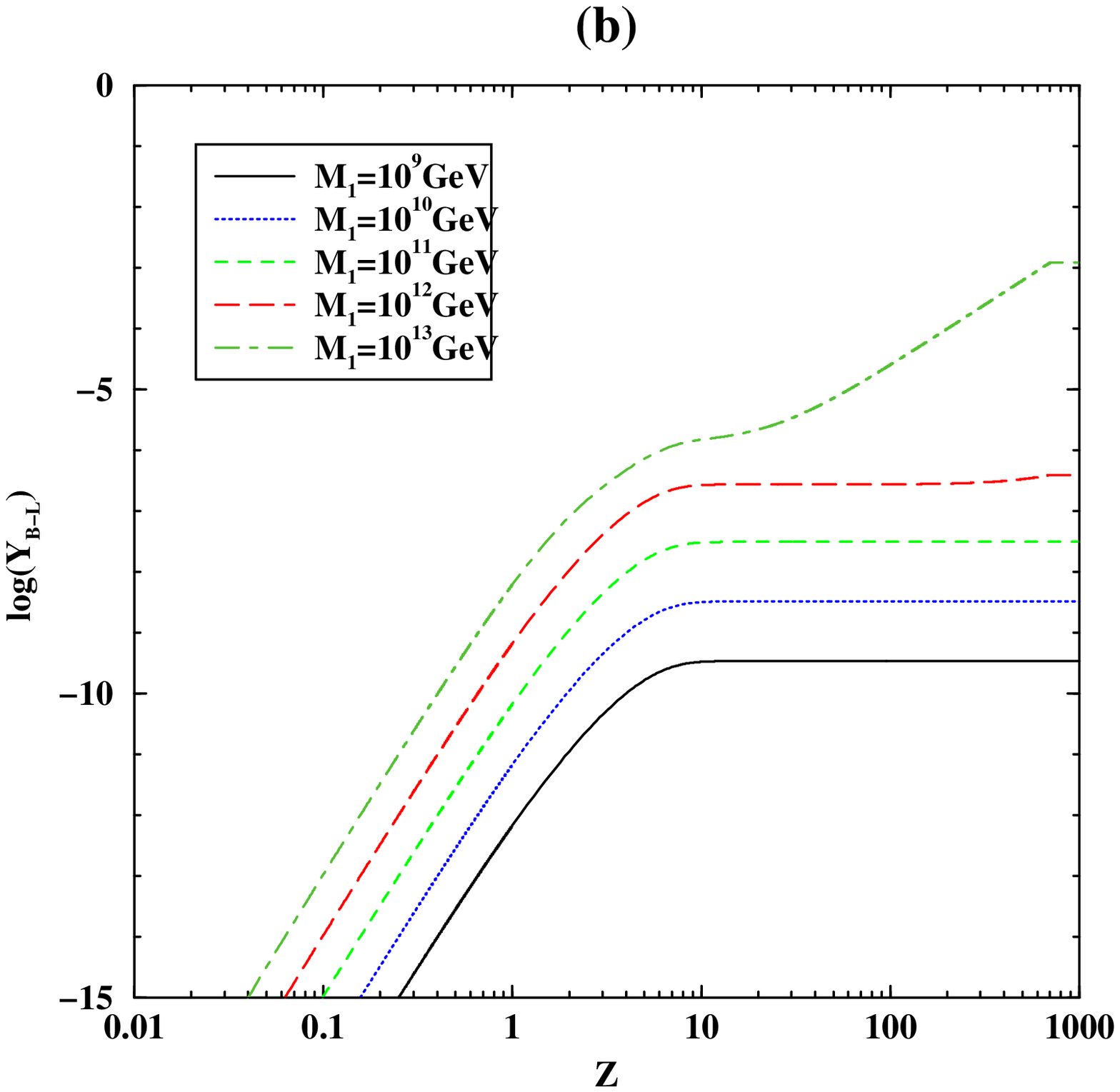, width=0.5\textwidth}
\caption{Evolution of the $B-L$ asymmetry for $\tilde{m}_1=10^{-4}\ev$,  
$\eta_{B-L}=10^{13}\gev$ and different values of $M_1$ (a) in absence
of cosmic strings, and (b) in presence of cosmic strings. The $CP$ 
violation parameter $\epsilon_1$ has been given its maximal value.}
\label{fig:eta=13-4}
\end{figure}

\begin{figure}[htbp]
\epsfig{file=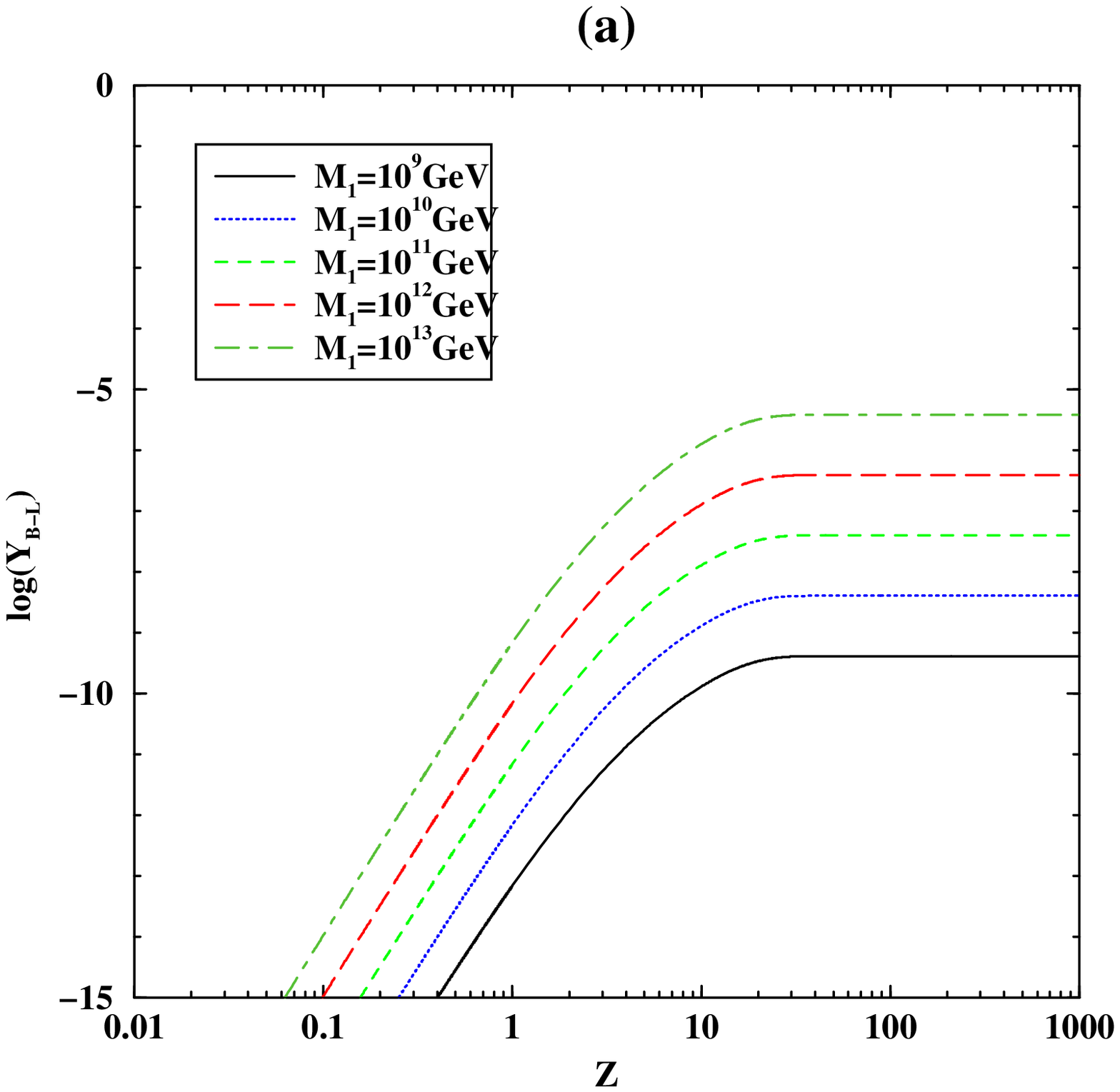, width=0.5\textwidth}
\epsfig{file=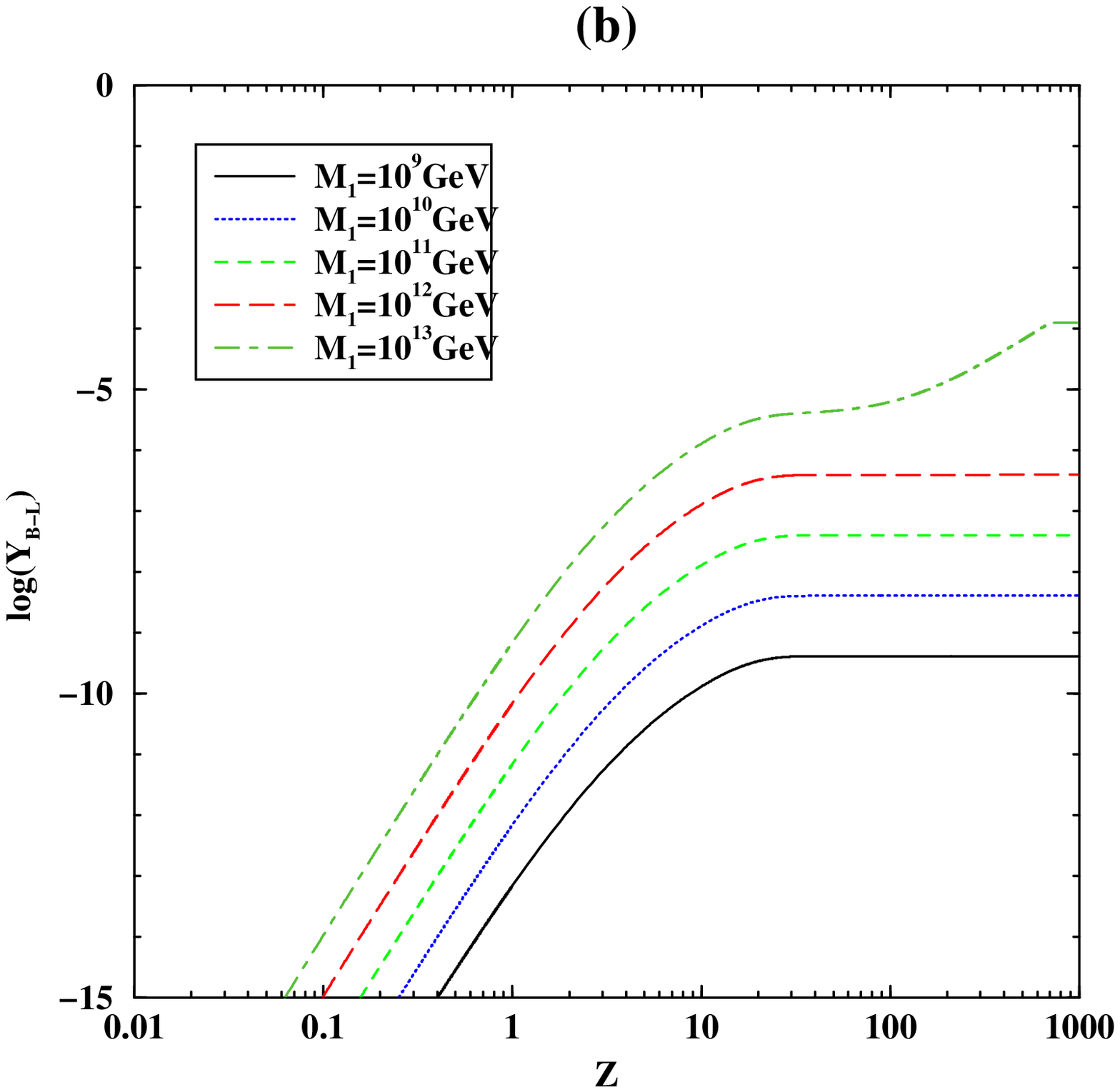, width=0.5\textwidth}
\caption{Evolution of the $B-L$ asymmetry for $\tilde{m}_1=10^{-5}\ev$,
$\eta_{B-L}=10^{13}\gev$ and different values of $M_1$ (a) in absence
of cosmic strings, and (b) in presence of cosmic strings. The $CP$
violation parameter $\epsilon_1$ has been given its maximal value.} 
\label{fig:eta=13-5}
\end{figure}

In the usual thermal scenario the $B-L$ asymmetry depends on 
$\tilde{m}_1$ and $M_1$. In presence of cosmic strings there is in 
addition an explicit dependence on $\eta_{B-L}$ since
the injection rate of $N_R$'s from the cosmic string loops explicitly 
depends on it. To illustrate the effect of the cosmic string generated 
non-thermal $N_R$'s on the evolution of the $B-L$ asymmetry, 
we display in Figures \ref{fig:eta=13-4} and \ref{fig:eta=13-5} the 
results obtained by numerically solving the Boltzmann 
equations described above for some specific values of the various relevant 
parameters. It can be seen from Figure~\ref{fig:eta=13-4}(a), for example,  
that in the absence of cosmic strings the $B-L$ asymmetry approaches the 
final value at $T\simeq 0.1 M_1$ when all the wash out processes fall 
out of equilibrium. In contrast, in the presence of cosmic strings, for 
sufficiently large values of $M_1$, the $B-L$ asymmetry 
continues to build up until the injection rate of $N_R$'s from cosmic 
string loops becomes insignificant. For $h_1=1$ this happens around 
$T\simeq M_1/600$ (see Figure \ref{fig:eta=13-4}(b)), which is far 
lower than in the purely thermal case. As a result of this, in presence 
of cosmic strings, for a fixed value of the symmetry breaking scale 
$\eta_{B-L}=10^{13}\gev$, for example, the final $B-L$ 
asymmetry is enhanced by three orders of magnitude for the effective 
neutrino mass $\tilde{m}_1=10^{-4}\ev$ (Fig.~\ref{fig:eta=13-4}(b)) and 
by two orders of magnitude for $\tilde{m}_1=10^{-5}\ev$ 
(Fig.~\ref{fig:eta=13-5}(b)). For $\eta_{B-L}=10^{13}\gev$,
the effect of cosmic string essentially disappears for $h_1<0.01$ 

The above results, in particular the dependence on the value of 
$\tilde{m}_1$, can be understood as follows: First let us consider the 
purely thermal leptogenesis case, i.e., in absence of cosmic strings. 
Below a certain temperature when the wash out processes fall out of 
equilibrium the asymmetry produced by the decay of $N_1$'s does not get 
wiped out, and the produced $B-L$ asymmetry remains as the final 
asymmetry. Since the decay rate of $N_1$'s depends 
linearly on $\tilde{m}_1$ (see equation (\ref{dilution})), 
a larger value of $\tilde{m}_1$ implies that the condition of 
decay in out-of-equilibrium situation is satisfied at a 
later time when the abundance of thermal $N_1$'s is smaller, 
thus yielding a smaller final value of $Y_{B-L}$. And as expected, this 
effect is larger for larger values of $M_1$. Now, in 
the presence of cosmic strings, for values 
of $\eta_{B-L}$ for which the contribution of the cosmic string 
generated $N_1$'s to $Y_{B-L}$ always remains insignificant compared 
to that due to the thermal $N_1$'s, the dependence of the final 
value of $Y_{B-L}$ on $\tilde{m}_1$ is 
essentially the same as in the absence of strings as explained above. 
However, for those values of $\eta_{B-L}$ for which the string 
contributions to $Y_{B-L}$ are significant compared to the thermal 
contribution, the dependence of the final value of $Y_{B-L}$ on 
$\tilde{m}_1$ is opposite to that in the absence of cosmic strings, i.e.,  
the string contribution increases with increasing values of $\tilde{m}_1$. 
This is easy to understand: The string generated $N_R$'s make dominant 
contribution to $Y_{B-L}$, if at all, only at late times ($T\ll M_1$) when 
the abundance of the thermal $N_R$'s has fallen to insignificant levels. 
The decay of the string generated $N_R$'s at such late times 
automatically satisfies the out-of-equilibrium condition. In this  
situation a larger value of $\tilde{m}_1$ simply implies  
a larger rate of decay of the $N_R$'s leading to a quicker development of 
the string contribution to $Y_{B-L}$. This, together with the fact that 
the injection rate of the $N_R$'s from cosmic string loops is higher at 
earlier times, leads to a higher final value of $Y_{B-L}$. Figures 
\ref{fig:eta=13-4}(b) and \ref{fig:eta=13-5}(b) plotted for 
$\tilde{m}_1=10^{-4}\ev$ and $\tilde{m}_1=10^{-5}\ev$ simultaneously bear 
out these expectations. 

Note that in Figures \ref{fig:eta=13-4} and \ref{fig:eta=13-5} the CP 
asymmetry parameter $\epsilon_1$ has been taken to have its maximum  
value, i.e., $\sin\delta=1$ in equation (\ref{epsilon_1}). Clearly, for 
this value of $\epsilon_1$ the produced $B$ asymmetry for 
$\eta_{B-L}=10^{13}\gev$ and $M_1\gsim 10^{10}\gev$ (i.e., 
$h_1\gsim10^{-3}$) exceeds the observed value given by 
$Y_{B-L}^{\rm obs}\sim O(10^{-10})$ even for purely thermal leptogenesis.  
Including the effect of cosmic strings only makes the situation worse. 
This implies that we must place an upper bound on the CP asymmetry 
phase $\delta$ depending on the values of $\eta_{B-L}$, $M_1$ and 
$\tilde{m}_1$, in order not to overproduce the $B$ asymmetry. 
At the same time, to produce the observed BAU for 
$\eta_{B-L}=10^{13}\gev$ and $h_1=1$, for example, we require 
$\sin\delta=\sin\delta_{\rm reqd}=O(10^{-7})$ and  
$\sin\delta_{\rm reqd}=O(10^{-6})$ for $\tilde{m}_1=10^{-4}\ev$
and $\tilde{m}_1=10^{-5}\ev$, respectively. These values of $\sin\delta$ 
diminish the purely thermal contributions, $Y_{B-L}^{\rm 
th}$, to $O(10^{-13})$ and $O(10^{-12})$, respectively.

We have calculated the value of $Y_{B-L}$ for a range of 
values of $\eta_{B-L}$, $h_1$ and $\tilde{m}_1$, both in the purely 
thermal case, $Y_{B-L}^{\rm th}$, and with the effect of cosmic strings 
included, $Y_{B-L}^{\rm th+st}$. The results for 
$\eta_{B-L}=10^{12}\gev$ and $\eta_{B-L}=10^{11}\gev$ are 
summarized in Tables \ref{table-1} and \ref{table-2}, respectively, where 
we also indicate the order of magnitude of $\sin\delta_{\rm reqd}$, the 
value 
of $\sin\delta$ required to produce the observed $B$ asymmetry, for 
each set of the chosen parameters. 

From these Tables as well as from Figures \ref{fig:eta=13-4} and 
\ref{fig:eta=13-5} we see that there exist ranges of values of the 
parameters $\eta_{B-L}$, $h_1$, $\tilde{m}_1$ and $\delta$ for which the 
cosmic string generated $N_R$'s make the dominant contribution to and 
indeed produce the observed BAU while the purely thermal leptogenesis 
mechanism is not sufficient.  

\begin{table}[htbp]
\begin{center}
\begin{tabular}{|c|c|c|c|c|c|c|}
\hline
$h_1$ & \multicolumn{3}{|c|} {$\tilde{m}_1=10^{-4}\ev$} & 
\multicolumn{3}{|c|} {$\tilde{m}_1=10^{-5}\ev$} \\ \cline{2-7}
 & $Y^{\rm th}_{B-L}$ & $Y^{\rm th+st}_{B-L}$ & $\log(\sin\delta_{\rm 
reqd})$
& $Y^{\rm th}_{B-L}$ & $Y^{\rm th+st}_{B-L}$ & $\log(\sin\delta_{\rm 
reqd})$\\
[2mm]\hline
1 & $2.72\times 10^{-7}$ & $1.23\times 10^{-5}$ & $-5$ & 
$3.90\times 10^{-7}$ & $1.59\times 10^{-6}$ & $-4$\\ [2mm]\hline
0.1 & $3.10\times 10^{-8}$ & $3.22\times 10^{-8}$ & $-2$ 
& $3.97\times 10^{-8}$ & $3.98\times 10^{-8}$ & $-2$\\ 
[2mm]\hline
0.01 & $3.25\times 10^{-9}$ & $3.25\times 10^{-9}$  & $-1$ 
& $4.04\times 10^{-9}$ & $4.04\times 10^{-9}$ & $-1$\\ [2mm]\hline
0.001 & $3.39\times 10^{-10}$ & $3.39\times 10^{-10}$ & 0 & 
$4.11\times 10^{-10}$ & $4.11\times 10^{-10}$ & 0\\[2mm]\hline
\end{tabular}
\end{center}
\caption{The $B-L$ asymmetry in the purely thermal leptogenesis 
case ($Y^{\rm th}_{B-L}$) and in presence of $B-L$ cosmic strings 
($Y^{\rm th+st}_{B-L}$) for $\eta_{B-L}=10^{12}\gev$ and different values 
of the Yukawa coupling $h_1$ and the effective neutrino mass 
$\tilde{m}_1$. The order of magnitude of the value of $\sin\delta$ 
required to 
produce the observed BAU, $\sin\delta_{\rm reqd}$, for each set of 
values of the parameters $\tilde{m}_1$ and $h_1$, is also given.} 
\label{table-1}
\end{table}

\begin{table}[htbp]
\begin{center}
\begin{tabular}{|c|c|c|c|c|c|c|c|}
\hline
$h_1$ & \multicolumn{3}{|c|} {$\tilde{m}_1=10^{-4}\ev$} &
\multicolumn{3} {|c|} {$\tilde{m}_1=10^{-5}\ev$}\\ \cline{2-7}
 & $Y^{\rm th}_{B-L}$ & $Y^{\rm th+st}_{B-L}$ & $\log(\sin\delta_{\rm 
reqd})$
& $Y^{\rm th}_{B-L}$ & $Y^{\rm th+st}_{B-L}$ & $\log(\sin\delta_{\rm 
reqd})$\\
[2mm]\hline
1 & $3.10\times 10^{-8}$ & $1.51 \times 10^{-7}$ & $-3$ & $3.97\times
10^{-8}$ & $5.17\times 10^{-8}$ & $-2$ \\ [2mm]\hline
0.1 & $3.25\times 10^{-9}$ & $3.27\times 10^{-9}$ & $-1$
& $4.04\times 10^{-8}$ & $4.04\times 10^{-9}$ & $-1$\\
[2mm]\hline
0.01 & $3.39\times 10^{-10}$ & $3.39\times 10^{-10}$ & 0 & $ 4.11\times
10^{-10}$ & $4.11\times 10^{-10}$ &  0\\ [2mm]\hline
\end{tabular}
\end{center}
\caption{Same as Table \ref{table-1} but for $\eta_{B-L}=10^{11}\gev$.} 
\label{table-2}
\end{table}
\section{Conclusion\label{sec:conclusion}}
We have studied the effect of $B-L$ cosmic strings arising from the 
breaking of a $U(1)_{B-L}$ gauge symmetry, on the baryon asymmetry of the 
Universe. The disappearance of closed loops of $B-L$ cosmic strings can 
produce heavy right handed neutrinos, $N_R$'s, whose $CP$-asymmetric decay 
in out-of-thermal equilibrium condition can give rise to a net lepton 
($L$) asymmetry which is then converted, due to sphaleron transitions, to 
a Baryon ($B$) asymmetry. We have solved the relevant Boltzmann 
equations that include the effects of both thermal and string generated 
non-thermal $N_R$'s. By exploring the parameter region spanned by the 
effective light neutrino mass parameter $\tilde{m}_1$, the mass $M_1$ of 
the lightest of the heavy right-handed neutrinos (or equivalently the 
Yukawa coupling $h_1$) and the scale of the $B-L$ symmetry breaking, 
$\eta_{B-L}$, we found that there exist ranges of values of 
these parameters, in particular with $\eta_{B-L} > 10^{11}\gev$ 
and $h_1\gsim 0.01$, for which the cosmic string generated non-thermal 
$N_R$'s can give the dominant contribution to, and indeed produce, the 
observed Baryon Asymmetry of the Universe when the purely thermal 
leptogenesis mechanism is not sufficient. We have also discussed how, 
depending on the values of $\eta_{B-L}$, $\tilde{m}_1$ and $h_1$, our 
results lead to upper bounds on the $CP$ violating phase $\delta$ that 
determines the relevant $CP$ asymmetry in the decay of the heavy right 
handed neutrino responsible for generating the $L$-asymmetry. 

\section*{Acknowledgment}
The work of UAY and NS is supported in part by a grant from
Department of Science and Technology. UAY wishes to thank the 
hospitality  of PRL Ahmedabad where part of this paper was completed.

\end{document}